\documentstyle[12pt]{article}
\newtheorem{theorem}{Theorem}
\newtheorem{itlemma}{Lemma}[section]
\newtheorem{itproposition}[itlemma]{Proposition}
\newtheorem{itcorollary}[itlemma]{Corollary}
\newtheorem{itremark}[itlemma]{Remark}
\newtheorem{itremarks}[itlemma]{Remarks}
\newtheorem{itdefinition}[itlemma]{Definition}
\newtheorem{itexample}[itlemma]{Example}

\newenvironment{lemma}{\begin{itlemma}\rm}{\end{itlemma}} 
\newenvironment{remark}{\begin{itremark}\rm}{\end{itremark}} 
\newenvironment{remarks}{\begin{itremarks} \rm}{\end{itremarks}}
\newenvironment{corollary}{\begin{itcorollary}\rm}{\end{itcorollary}}
\newenvironment{proposition}{\begin{itproposition}\rm}{\end
{itproposition}}
\newenvironment{definition}{\begin{itdefinition}\rm}{\end{itdefinition}}
\newenvironment{example}{\begin{itexample}\rm}{\end{itexample}}
\newenvironment{fact}{\noindent {\em Fact}. \ \ }{\hfill \medskip}
\newenvironment{proof}{\noindent {\em Proof}.\ \
}{\hspace*{\fill}$\Box$\medskip}
\newenvironment{claim}{\noindent {\em Claim}. \ \ }{\hfill \medskip}
\newcommand{\be}[1]{\begin{equation}\label{#1}}
\newcommand{\ee}{\end{equation}}
\newcommand{\bl}[1]{\begin{lemma}\label{#1}}
\newcommand{\br}[1]{\begin{remark}\label{#1}}
\newcommand{\brs}[1]{\begin{remarks}\label{#1}}
\newcommand{\bt}[1]{\begin{theorem}\label{#1}}
\newcommand{\bd}[1]{\begin{definition}\label{#1}}
\newcommand{\bp}[1]{\begin{proposition}\label{#1}}
\newcommand{\bc}[1]{\begin{corollary}\label{#1}}
\newcommand{\bfact}[1]{\begin{fact}\label{#1}}
\newcommand{\bex}[1]{\begin{example}\label{#1}}
\newcommand{\ec}{\end{corollary}}
\newcommand{\efact}{\end{fact}}
\newcommand{\eex}{\end{example}}
\newcommand{\el}{\end{lemma}}
\newcommand{\er}{\end{remark}}
\newcommand{\ers}{\end{remarks}}
\newcommand{\et}{\end{theorem}}
\newcommand{\ed}{\end{definition}}
\newcommand{\ep}{\end{proposition}}
\newcommand{\epr}{\end{proof}}
\newcommand{\bpr}{\begin{proof}}
\newcommand{\bcl}{\begin{claim}}
\newcommand{\ecl}{\end{claim}}

\newcommand{\bi}{\begin{itemize}}
\newcommand{\ei}{\end{itemize}}
\newcommand{\ben}{\begin{enumerate}}
\newcommand{\een}{\end{enumerate}}
\newcommand{\text}[1]{\hbox{\rm \ #1\ \/}}

\newcommand{\vs}{\vspace{0.25cm}}

\newcommand{\r}{{{\cal R}}}

\begin{document}

\begin{center}
{\Large {Model Identification
for  Spin Networks\\ }}
\end{center}

\bigskip

\begin{center}

{Francesca Albertini\\ \vs Dipartimento di Matematica Pura ed
Applicata,\\ Universit\`a di Padova,\\ via Belzoni 7,\\ 35100
Padova, Italy.\\ Tel. (+39) 049 827 5966\\ email:
albertin@math.unipd.it}\\

\vs

  {Domenico D'Alessandro \\ \vs Department of Mathematics\\ Iowa
State University \\ Ames, IA 50011,  USA\\ Tel. (+1) 515 294
8130\\ email: daless@iastate.edu}

\end{center}

\begin{abstract}
We consider the problem of determining the unknown parameters of the 
Hamiltonian of a network of spin $\frac{1}{2}$ particles. In particular,
we study  experiments in which the system is driven by an
externally applied electro-magnetic field and the expectation value of
the total magnetization is measured.  
Under appropriate assumptions, 
we
prove that,  if it is possible to prepare 
the system in a known  initial state,  the 
above experiment  allows  to identify
the parameters  of the Hamiltonian. In the case where the initial state is
itself an unknown parameter,  we characterize all the pairs
Hamiltonian-Initial State  
 which give the same value of
the magnetization for every form of the driving electro-magnetic
field. The analysis  is motivated by recent results
on the isospectrality of Hamiltonians describing Magnetic Molecules.  

\end{abstract}

\section{Introduction}
\label{intro}
In recent years, chemists have developed 
methods to synthesize large organometallic 
molecules which contain a core of magnetic transition
metal ions  interacting via electronic superexchange
interactions \cite{Ash}, \cite{Barbara}, \cite{MRS}.  One of the main
advantages of this technology is that it makes it  possible to arrange the
molecules in regular van der Waals crystals in which the magnetic
interactions among different molecules are negligible. Thus, every
cluster of this kind behaves  like  an assembly of {\it identical} and
{\it independent}  
nanosize magnets, each corresponding to one molecule. For this reason, 
these novel systems are now deemed ideal to study {\it fundamental questions}
concerning magnetism at the molecular level. In fact, even the simplest
of these systems displays several new classical and quantum mechanical
phenomena. One example is {\it macroscopic quantum tunneling of
 magnetization}  \cite{trentadue}, \cite{trentaduebis}, 
a fascinating issue of relevance to a
variety of mesoscopic systems and 
nanostructures. This paper is a study on the determination
of the parameters for Hamiltonians describing these {\it magnetic molecules}. 
The determination of exchange constants for these molecules  has been
traditionally obtained by measuring the temperature-dependent magnetic
susceptibility  or other thermodynamic properties of these
compounds. This technique relies on
the assumption that there is a one
to one correspondence between the
spectrum of the Hamiltonian of the
system in a static magnetic field,  and its thermodynamic properties as
well as the numerical values of the model parameters. However, a recent
article \cite{luban} has shown that  quantum Heisenberg spin systems,  such
as magnetic molecules,  may have different coupling parameters but the same
energy spectrum and hence the same thermodynamic properties.  This raises
the question of whether a {\it dynamic} technique could be used to identify
the parameters of a spin network. In particular, as in NMR or EPR
experiments, one could try to
let the system evolve under the action of a driving time varying 
electro-magnetic
field, measure the total magnetization and,  from the measured
value,  infer the value of the parameters of the system. The question of
whether these experiments are able to identify the unknown parameters 
can be tackled at 
different
levels according to how much we assume known about the system under
investigation; whether, for example, we assume the system 
prepared in a known initial state or 
the initial state itself a parameter to be identified.

\vs


\vs

In this paper, we consider networks of spin $\frac{1}{2}$ particles and 
present a positive answer to the above question. We prove 
that systems which give the same input-output
behavior for any given state are the same. This shows that, if we can
opportunely prepare an initial state we can use the above scheme to
identify the parameters of the system exactly. If the 
 initial state of the system is unknown we prove t
hat there are only two possibilities (up to
permutations of the spins) for two pairs
Hamiltonian-Initial State to give the same input output behavior. They are
either the same or  the exchange constants have opposite signs and the
initial states are related in a way we shall describe. So, in this case,
the given experiments identify one of two possible systems giving the
observed behavior.

\vs

The paper is organized as follows. In the next
section, we give the basic definitions and state the 
problem we want to solve in mathematical terms. In Section 3, we give the
two results above described and prove  the first one. 
This
gives us the opportunity to elaborate on the control theoretic 
concepts of {\it observability}
and {\it controllability} and their role in the parameter identification
problem. The proof of the second result is much longer  and it is
presented in Sections 4 and 5. Section 6 presents some conclusions
and a discussion of the results. 

\section{Definitions and Statement of the Problem}
\label{basic}

  We consider an Heisenberg spin Hamiltonian of the form
  \be{dinamica}
H(t):= i(A+B_xu_x(t)+B_yu_y(t)+B_zu_z(t)), \ee with \be{dinamica1}
\begin{array}{ccl}
A &:= & -i \sum_{k<l, k,l=1}^n J_{kl}(I_{kx,lx}+ I_{ky,ly}+
I_{kz,lz}), \\
   & & \\
B_{v}&:= & -i (\sum_{k=1}^n \gamma_k I_{kv}), \ \ \text{ for }
v=x,\, y, \text{ or } z. \end{array} \ee
  For a network of $n$ spin
$\frac{1}{2}$ particles, the matrix $I_{k_1v_1,...,k_rv_r}$ is the
Kronecker product of $n$ matrices equal to the $2 \times 2$
identity except in the $k_j-$th ($j=1,...,r$) position(s) occupied by
the Pauli matrix $\sigma_{v_j}$, $v_j=x,y,z$. Recall (see e.g.
\cite{sakurai}) that the Pauli matrices
are defined as
\be{Pauli} \sigma_x:= \frac{1}{2} \pmatrix{0 & 1
\cr 1 & 0},  \qquad \sigma_y:=\frac{1}{2} \pmatrix{0 & -i \cr i & 0}, 
\qquad \sigma_z:=\frac{1}{2}\pmatrix{1 & 0 \cr 0 & -1}. \ee

\vs

We denote by $\cal R$ the set of possible values for the density
matrix, i.e. Hermitian, positive semidefinite matrices
with trace one. Let $\rho(t,u_x,u_y,u_z,\rho_0)$, the density matrix
solution of the Schr\"odinger equation corresponding to the controls
$u_x(t), \,
u_y(t), \, u_z(t)$ and initial  condition $\rho_0$. We assume
that it is possible to observe the expectation
value of the total magnetization in the
$x$, $y$, and $z$ direction, namely:
\be{output}
M_{v}(t)
:=M_v(t;\rho_0,u_x,u_y,u_z)
:= Tr(S^{TOT}_{v}\rho(t,u_x,u_y,u_z, \rho_0)),
  \ee where
\[S^{TOT}_{v}=\sum_{k=1}^{n}I_{kv}, \ \ \text{ for } v=x,y,z.\]
We study the possibility of distinguishing the parameters by a
single measurement of one of the above outputs. More precisely, we
denote by $\Sigma \equiv \Sigma (n,J_{kl},\gamma_k)$ a model
described by the equations (\ref{dinamica}) and (\ref{dinamica1}),
and by $(\Sigma,\rho_0)\equiv(\Sigma (n, J_{kl}, \gamma_k),\rho_0)$
a model given with fixed initial state $\rho_0$.
Thus, for given control functions $u_x$, $u_y$, and $u_z$,
$M_v(t;\rho_0,u_x,u_y,u_z)$, for $v=x,\, y$, and $z$, are the
corresponding output functions.
The parameters $J_{kl}$ and
$\gamma_k$ along with the number $n$ of spins (and the value of the
initial state $\rho_0$) characterize the
model. The question of parameter identifiability through a single
measurement of the magnetization can be posed by identifying the
models (or set of parameters) that give the same input-output
behavior. We have the following definition.

\bd{equivalenti}
Consider two models $\Sigma$ and $\Sigma'$.   We mark with a prime
$'$ all the symbols concerning system $\Sigma'$.
\begin{itemize}
    \item $\Sigma$ and $\Sigma'$
are {\it equivalent} and we write
$$
\Sigma \sim \Sigma'
$$ if and only if $n=n'$ and for any given common
initial condition $\rho_0$ and control functions $u_x,u_y,u_z$,
  we have
\[M_v(t;\rho_0,u_x,u_y,u_z)\, =\, M'_v(t;\rho_0,u_x,u_y,u_z), \
\ \text{ for } v=x,\, y,\, z.\]
\item
Two pairs model-initial state
$(\Sigma, \rho_0)$ and $(\Sigma',\rho_0')$
are equivalent and we write
\[
(\Sigma,\rho_0) \sim (\Sigma',\rho_0'),\] if for  all control
functions $u_x$, $u_y$, and $u_z$, we have
\[M_v(t;\rho_0,u_x,u_y,u_z)\, =\, M'_v(t;\rho_0',u_x,u_y,u_z), \
\ \text{ for } v=x,\, y,\, z.\]
\end{itemize}\ed

\vs

\bd{controllabilita}
A model is {\it controllable} if by varying opportunely the control
functions $u_x,u_y,u_z$, it is possible
to drive the evolution operator from the identity to any unitary matrix.
\ed

\vs

For general quantum systems controllability can be checked by verifying
the so-called Lie Algebra Rank Condition \cite{Tarn} which means
that the matrices $A$ and $B$'s characterizing the dynamics
(cfr.(\ref{dinamica})-(\ref{dinamica1}))
generate the whole Lie Algebra $su(\tilde n)$ (or $u(\tilde n)$) where
$\tilde n$ is the dimension of the underlying subspace ($2^n$ for the
case of networks of spin $\frac{1}{2}$'s). To a network of spin
$\frac{1}{2}$ one can associate a graph whose nodes 
represent the  particles  and an edge connects two nodes if and
only if the corresponding exchange 
constant is different from zero. In the case of
spin networks with different gyromagnetic ratios the system is
controllable if and only if the graph associated to the network  is
connected, and sufficient controllability conditions can be
given for the general case \cite{confraLAA}.

\vs

\bd{osservabilita}
A model is {\it observable} if there are no two different states which 
give the
same output for every set of control functions.
\ed

\vs

Observability can be checked by verifying that the {\it Observability 
Space}
${\cal V}$ is equal to $su(\tilde n)$ ($\tilde n$ again is the dimension
of the underlying Hilbert space) \cite{CDC2003}. Considered the matrix
that characterizes the output, $S_v^{TOT}$ in our case, the
observability space ${\cal V}$ is  the vector space spanned by the
matrices, \footnote{$ad_R^kT:=[R,[R,...[R,T]]]$ where the Lie
bracket is taken $k$ times.} \be{mat} ad_{B_{j_1}}^{k_1}
ad_{B_{j_2}}^{k_2} \cdot \cdot \cdot ad_{B_{j_r}}^{k_r}
iS^{TOT}_{v}  \ \ v=x,y,z, \ee where $j_{k}\in \{0,1,2,3\}$ and 
$B_{0}=A$, $B_{1}=B_{x}$,
$B_{2}=B_{y}$, and $B_{3}=B_{z}$. The following fact holds true 
\cite{CDC2003}.
\bl{CimpO}
Controllability implies observability.
\el

\vs

In parametric identification problems, it makes sense to restrict
ourselves to observable systems since the unobservable dynamics does not
contribute to the output which is our tool to identify the
system. Moreover,  we
want to check that
observable systems which give the same input-output behavior, namely
that are equivalents, have the same parameters. We can state the
following two problems.

\vs

\noindent {\bf Problem 1} Characterize the classes of observable
equivalent spin models.

\vs

\noindent {\bf Problem 2} Characterize the classes of observable
equivalent pairs model-initial state.

\vs

We shall see that the equivalence classes in Problem 1 consist of a
single element. We shall solve Problem $2$ restricting ourselves to
networks that have different gyromagnetic ratios (or for which the spins
can be selectively addressed) and controllable.

\section{Main Results}
\label{Main}

In the following, we shall always denote by $\rho(t)$ and $\rho'(t)$ two
trajectories corresponding to the same controls $u_x,u_y,u_z$ for the
models $\Sigma$, $\Sigma'$, respectively. The following Proposition
whose proof we relegate to Appendix A, will be used used in
the proof of both our main results. We notice here that the proof
although presented for the case of spin Heisenberg systems can be
adapted to any bilinear finite dimensional quantum control system.

\vs

\bp{prop1} Let  $(\Sigma, \rho_0)$  and $(\Sigma',\rho_0')$ be the
two fixed pairs. Then, the following are equivalent:
\begin{itemize}
\item[(a)] $(\Sigma,\rho_0)\sim (\Sigma',\rho_0')$,
\item[(b)] For all control functions $u_x(t)$, $u_y(t)$, and
$u_z(t)$, we
have:
\be{tra} Tr(F\rho(t)) =
Tr(F'\rho'(t)), \ee for all $F\in \cal V$ and $F' \in {\cal V}'$,
with $F'$ constructed as $F$ changing all $B_i$ in $B'_i$ (see
(\ref{mat})).
\end{itemize}
\ep

\vs

We now state the first of our two main results.

\vs

\bt{main1} Let $\Sigma(n,J_{kl},\gamma_k):= \{ A, B_x,B_y,B_z 
\}$ and   
$\Sigma'(n,J_{kl}',\gamma_k'):=\{A',  B_x',B_y',B_z'\}$ 
be two equivalent models. Assume one of them is observable. Then $A=A'$,
$B_{x,y,z}=B_{x,y,z}'$.
\et

\bpr From the equivalence of the models
and specializing (\ref{tra}) of \ref{prop1}
to a common initial condition,
we obtain
\be{traplus}
Tr(F\rho_0)=Tr(F'\rho_0),
\ee
for every $\rho_0 \in \cal R$. Therefore $F=F'$. From the observability
assumption $F$ (and $F'$)  span all of $su(2^n)$ ($n$ here is the number
of spins, which is assumed to be the same). Since $F$ is a generic 
element of $\cal V$, we have
$[A,F]=[A',F']=[A',F]$ and therefore
\be{jhlopl}
[A-A',F]=0.
\ee
Since $F$ spans all of $su(2^n)$, $A-A'$ must be zero. Analogously one
can prove $B_{x,y,z}=B_{x,y,z}'$.

\epr

\vs

Notice that,  although presented for the case of spin networks,  Theorem
\ref{main1}
holds for any finite dimensional quantum system and essentially says
that two observable models with the same input-output behavior, for a
every  state,  must be
equal. The assumption of observability can be checked by checking
controllability and applying Lemma \ref{CimpO}. Conditions for
controllability of spin networks are given in \cite{confraLAA}.

\vs

We now consider a more difficult problem since we assume to have much
less knowledge of the model to be identified. We assume not to know its
dimension nor the initial condition. We perform black-box type of
experiments on two pairs model-initial state and we obtain the same
results. We investigate what can be said about the two models. We assume
that all the  gyromagnetic
ratios $\gamma_k$, ($\gamma_k'$) are different. This fact implies that the
(mild) assumption that the graph
associated to the spin network is connected is equivalent to 
controllability \cite{confraLAA}. We shall assume this. 
Under this assumption, we 
can easily rule  out the case
in which the responses of the systems are both identically zero. In
this case (and only in this case) the corresponding initial density
matrices are scalar matrices and nothing more can be said about two
equivalent models. So we will assume that the two initial states are
not scalar matrices.

Before stating the result, we need to introduce some more notation.
We denote by ${\cal I}_o$ (${\cal I}_e$) the subspace of the
Hermitian matrices of dimension $2^n$ generated by Kronecker
products that contain an odd (even) number of Pauli matrices (and
the rest $2 \times 2$ identity matrices).
Moreover, if
   $\pi$ is a permutation of the set $\{1,\ldots,n\}$, we denote
by $P_\pi$  the  matrix which transforms Kronecker products of $n$
$2 \times 2$ matrices according to the permutation $\pi$ (cfr.
\cite{HJTop} pg. 260), namely for every $n-$ple of $ 2 \times 2$
matrices $K_1,...,K_n$ we have \be{transfo} P_\pi (K_1 \otimes K_2
\otimes \cdot \cdot \cdot \otimes K_n) P_\pi=
  K_{\pi(1)} \otimes K_{\pi(2)} \otimes \cdot \cdot \cdot \otimes
K_{\pi(n)}. \ee

\bt{main}
  Let
$(\Sigma,\rho_0) \equiv (\Sigma (n,J_{kl},\gamma_k),\rho_0)$ and
$(\Sigma',\rho'_0) \equiv(\Sigma' (n',J'_{kl},\gamma'_k),\rho_0')$
be two fixed models whose dynamics and output are given by
equations (\ref{dinamica}), (\ref{dinamica1}), and
(\ref{output}). Assume that both models are controllable, that all
the $\gamma_k$ and $\gamma_k'$ different from each other, and that
$\rho_0$ and $\rho_0'$ are not scalar matrices. Then the following are
equivalent:
\begin{itemize}
\item[(a)] $(\Sigma,\rho_0) \sim (\Sigma',\rho_0')$,
\item[(b)]  $n=n'$ and there exists a permutation $\pi$ of the
set $\{1,\ldots,n\}$ such that
\begin{enumerate}
\item $\gamma_k=\gamma'_{\pi(k)}$,
\item
denoting by
$\pi_{lk}^1=\min \{ \pi(l), \pi(k)\}$, and $\pi_{lk}^2=\max \{
\pi(l), \pi(k)\}$, for $1\leq l <k\leq n$,
then either:
\be{uguali}
\left\{ \begin{array}{l}
J_{lk}=J'_{\pi_{lm}^1\pi_{lm}^2}  \ \ \forall 1\leq l<k\leq n,\\
P_{\pi}\rho_0'P_{\pi}=\rho_{0};
\end{array} \right.
\ee or \be{opposti} \left\{ \begin{array}{l}
J_{lk}=-J'_{\pi_{lm}^1\pi_{lm}^2}  \ \ \forall 1\leq l<k\leq n,\\
\rho_{1}= \rho_{1}'\  \text{ and }\  \rho_2=-\rho_{2}';
\end{array} \right.
\ee
where $\rho_1$ and
$\rho_2$ (resp. $\rho_1'$ and $\rho_2'$)  are the components
of $\rho_0$ (resp. $P_{\pi}\rho_0'P_{\pi}$) in ${\cal I}_o$, ${\cal 
I}_e$,
respectively.
\end{enumerate}
\end{itemize}
\et

Equations (\ref{uguali}),  (\ref{opposti})
say that,  up to a permutation of the spins,  the exchange constants are
all the same or all opposite. In one case the initial conditions are the
same in the other case the components in ${\cal I}_o$ are   the same while
the components in ${\cal I}_e$ are opposite. The next two sections are devoted to the
proof of Theorem \ref{main}. We recall that we have two standing
assumptions in all the treatment. The models we are dealing with have
different $\gamma$'s and are controllable.

\section{Preliminary Results}
\label{preli}

In order to prove the implication $(a)  \Rightarrow (b)$ we shall
need some properties of equivalent pairs. We present them in this
section with all the proofs in Appendix B.
The following proposition says that equivalent pairs
(Model-Initial State) must have the same dimension.

\bp{prop2} Let  $(\Sigma, \rho_0)$  and $(\Sigma',\rho_0')$ be the
two fixed models. If they are equivalent, then $n=n'$ and there
exists a permutation $\pi$ of the set $\{1,\ldots,n\}$ such that:
\begin{enumerate}
\item $\gamma_k\, =\, \gamma'_{\pi(k)}$ for all $k\in\{1,\ldots,n\}$,
\item $Tr(I_{kv}\rho(t))=Tr(I_{\pi(k)v}\rho'(t))$
for all $t\geq 0$, all $k\in\{1,\ldots,n\}$, all $v\in \{x,y,z\}$,
and all possible trajectories $\rho(t)$ of  $(\Sigma, \rho_0)$  and
corresponding $\rho'(t)$ of  $(\Sigma',\rho_0')$.
\end{enumerate}
\ep

The proof of the next lemma
is not presented in the Appendix
  since it is just a notational
modification of the proof of Proposition \ref{prop1}.
\bl{lemma1}
Let $W$ and $W'$ be two Hermitian matrices of dimensions $2^n$ and
$2^{n'}$, respectively. If, for every trajectory $\rho(t)$ of
$(\Sigma, \rho_0)$  and corresponding trajectory $\rho'(t)$ of
$(\Sigma',\rho_0')$ we have \be{C1} Tr(W\rho(t))=Tr(W'\rho'(t)),
\ee then for every $F$, $F:=ad_{B_{j_1}}ad_{B_{j_2}}\cdot \cdot
\cdot ad_{B_{j_r}}W$, and corresponding $F'$,
$F':=ad_{B'_{j_1}}ad_{B'_{j_2}}\cdot \cdot \cdot ad_{B'_{j_r}}W'$,
with $r\geq 0$ and  $j_1,...,j_r\in \{0,1,2,3\}$, we also have
\be{C2} Tr(F\rho(t))=Tr(F'\rho'(t)). \ee \el \bl{lemma2} Let $W$
and $W'$ be two Hermitian matrices of dimensions $2^n$ and $2^{n'}$,
respectively. If, for every trajectory $\rho(t)$ of $(\Sigma,
\rho_0)$  and corresponding trajectory $\rho'(t)$ of
$(\Sigma',\rho_0')$ we have \be{C3} Tr(W\rho(t))=Tr(W'\rho'(t)),
\ee then,  for every $K$, $K:=ad_{B_{j_1}}ad_{B_{j_2}}\cdot \cdot
\cdot ad_{B_{j_r}}B_{j_{0}}$, and corresponding $K'$,
$K':=ad_{B'_{j_1}}ad_{B'_{j_2}}\cdot \cdot \cdot
ad_{B'_{j_r}}B'_{j_{0}}$, with $r\geq 0$ and  $j_0,...,j_r\in
\{0,1,2,3\}$, we also have \be{C4}
Tr([W,K]\rho(t))=Tr([W',K']\rho'(t)). \ee \el

\bl{lemma4} Let  $(\Sigma, \rho_0)$  and $(\Sigma',\rho_0')$ be
two fixed models. Assume that they are equivalent and let
$\pi$ be the permutation given by Proposition \ref{prop2}. If $W$
and $W'$ are two given Hermitian  matrices such  \be{C10}
Tr(W\rho(t))=Tr(W'\rho'(t)), \ee for every pair of corresponding
trajectories $\rho(t)$ and $\rho'(t)$, then it also holds \be{C11}
Tr([W,I_{kv}]\rho(t))=Tr([W',I_{\pi{(k)}v}]\rho'(t)),  \ \ \forall
\  k\in\{1,\ldots,n\}.\ee \el

\section{Proof of Theorem \ref{main}}
\label{complet}

Let $(\Sigma, \rho_0)$  and $(\Sigma',\rho_0')$ be the two given 
equivalent models. Assume that both models are controllable and  that
all the $\gamma_k$ and $\gamma_k'$ are
different from each other. 
We already know that $n=n'$ and $1$ and $2$ of Proposition
\ref{prop2} hold. To simplify the 
notations, we assume, without loss of generality, that
we have performed a change of coordinates in the second model so
that the permutation $\pi$ of Proposition \ref{prop2} is the identity. 
Thus we can  write 
\be{p1} \gamma_k =  \gamma'_{k},
\ \ \forall \ k\in\{1,\ldots,n\},
\ee and \be{p2} Tr(I_{kv}\rho(t))
=Tr(I_{kv}\rho'(t)) \ \ \forall \
k\in\{1,\ldots,n\}. \ee Equations (\ref{uguali})
and (\ref{opposti}) now read as: \be{puguali} \left\{
\begin{array}{l} J_{lk}=J'_{lk} \ \ \forall 1\leq l<k\leq n,\\
\rho_0'=\rho_{0};
\end{array} \right.
\ee or \be{popposti} \left\{ \begin{array}{l} J_{lk}=-J'_{lk}  \ \
\forall 1\leq l<k\leq n,\\ \rho_{1}= \rho_{1}' \ \text{ and } \
\rho_2= -\rho_{2}';
\end{array} \right.
\ee
where
  $\rho_1$ and $\rho_2$ (resp. $\rho_1'$ and $\rho_2'$)  are the 
components
of $\rho_0$ (resp. $\rho_0'$) in ${\cal I}_o$, ${\cal I}_e$,
respectively. We shall need the following two  lemmas whose proofs are
presented in Appendix C.

  \bl{plemma1} Assume that for all
$t\geq 0$, all possible trajectories $\rho(t)$ of  $(\Sigma,
\rho_0)$  and corresponding $\rho'(t)$ of  $(\Sigma',\rho_0')$, for
fixed values $1\leq k_{1},\ldots, k_{r}\leq n$, and fixed  $v_{j}\in
\{x,y,z\}$ we have: \be{p3} Tr\left(I_{k_1v_1,...,k_r v_r}
\rho(t)\right)=
  Tr\left(I_{k_1v_1,...,k_r v_r} \rho'(t)\right),
\ee
  Then:
  \begin{enumerate}
      \item
  equation (\ref{p3}) holds for any possible choice of the values of
  $v_{j}\in \{x,y,z\}$;
  \item
  \be{p4}
  Tr\left(\left[ \left[iI_{\bar l v_{\bar{l}}},[iI_{\bar k
  v_{\bar{k}}},A]\right], I_{k_1v_{1},...,k_rv_{r}}\right]
  \rho(t)\right)=
Tr\left(\left[ \left[iI_{\bar l v_{\bar{l}}},[iI_{\bar k
  v_{\bar{k}}},A]\right], I_{k_1v_{1},...,k_rv_{r}}\right]
  \rho'(t)\right),
\ee
  for every values $1\leq \bar{l}\neq \bar{k}\leq n$ and every
 $\{v_{\bar{l}}\neq v_{\bar{k}}\}\in \{x,y,z\}$.
  \end{enumerate}
\el

\bl{plemma2}
Assume that for all $t\geq 0$, all possible trajectories $\rho(t)$ of
$(\Sigma, \rho_0)$  and corresponding $\rho'(t)$ of
$(\Sigma',\rho_0')$,  for fixed values $1\leq
k_{1},\ldots,k_{r}\leq n$, $v_{j}\in \{x,y,z\}$ and for given
constants $\alpha$ and $\alpha'$, we have:
\be{p5}
\alpha Tr\left(I_{k_1v_1,...,k_r v_r} \rho(t)\right)= \alpha'
  Tr\left(I_{k_1v_1,...,k_r v_r} \rho'(t)\right),
\ee
  Then
  \begin{enumerate}
      \item
  For any pair of indices $\bar k, \bar l\in \{1,\ldots,n\}$ with $\bar k 
\in
\{k_1,...,k_r\}$   and $\bar l \notin
\{k_1,...,k_r\}$,
  \be{p6} \alpha J_{\bar k \bar l}
Tr\left(I_{k_1v_1,...,k_r v_r, \bar l \bar v} \rho(t)\right)= 
\alpha'J_{\bar k
\bar l}' Tr\left(I_{k_1v_1,...,k_r v_r,\bar l \bar v } \rho'(t)\right), 
\ee
  for any value $\bar v\in \{x,y,z\}$.
\item For any pair of   indices $\bar k, \bar l$ both in
$\{k_1,...,k_r\}$, (for example $\bar k=k_1$, $\bar l=k_2$)  then
\be{p7} \alpha J_{\bar k \bar l} Tr\left(I_{k_1v_1,k_3v_3,...,k_r
v_r} \rho(t)\right)= \alpha'J_{\bar k \bar l}'
Tr\left(I_{k_1v_1,k_3v_3,...,k_r v_r } \rho'(t)\right). \ee
\end{enumerate}
  \el

\subsection{$(a) \ \Rightarrow \ (b)$}

Fix any $1\leq k_1<k_2\leq n$, then, by applying statement $1.$ of
Lemma \ref{plemma2}, i.e. equation (\ref{p6}) with $\bar{k}=k_1$,
$\bar{l}=k_2$ to equation (\ref{p2}) with $k=k_1$, we have:
\be{p17} J_{k_1k_2} Tr\left( I_{k_1v_1,k_2v_2} \rho(t) \right) =
J'_{k_1k_2} Tr\left( I_{k_1v_1,k_2v_2} \rho'(t) \right), \ \  \
\forall\ v_1,v_2\in \{x,y,z\}. \ee Now, to the previous equality,
we apply statement $2.$ of Lemma \ref{plemma2}, i.e. equation
(\ref{p7}) with $\bar{k}=k_1$ and $\bar{l}=k_2$ to get:
\[
J_{k_1k_2}^2 Tr\left( I_{k_1v_1} \rho(t) \right) = J^{'2}_{k_1k_2}
Tr\left( I_{k_1v_1} \rho'(t) \right),
\]
which, by equation (\ref{p2}), implies: \be{p18}
J_{k_1k_2}^2=J_{k_1k_2}^{'2}. \ee 
Therefore the exchange constants are equal up to the sign. 
Now we prove, by the way of
contradiction that they are either all equal or all opposite i.e. 
 \be{p19} \left\{ J_{kl}=J'_{kl} \ \forall\,
1\leq k<l \leq n \right\} \text{ or } \left\{ J_{kl}=-J'_{kl} \
\forall\, 1\leq k<l \leq n. \right\} \ee Assume, by contradiction,
that (\ref{p19}) does not hold. By the  controllability
assumption and the results of \cite{confraLAA}
  we know that the graph associated to the network  is connected. From this fact it is not difficult to
see that if (\ref{p19}) is false, there must exist $3$ indices
$l$, $k_1$, and $k_2$ (here, to simplify notations, we assume
$1\leq l<k_1<k_2\leq n$, the other cases can be treated using
exactly the same arguments) such that: \be{p20}
J_{lk_1}=J_{lk_1}',\ \text{ and } \ J_{lk_2}=-J_{lk_2}'.\ee
  Using equation (\ref{p17}) we get:
\[
J_{lk_1}Tr ( I_{lv,k_1v_1} \rho(t)) \, =\, J'_{lk_1} Tr (
I_{lv,k_1v_1} \rho'(t)),\] for all $v,v_1\in \{x,y,z\}$ and all
corresponding  trajectories $\rho(t)$ and $\rho'(t)$. By applying
to the previous equality statement $1.$ of Lemma \ref{plemma2},
i.e.  equation (\ref{p6}) with $\bar{k}=l$ and $\bar{l}=k_2$, we
get:
  \be{p21}
  J_{lk_1} J_{lk_2}Tr ( I_{lv,k_1v_1,k_2v_2} \rho(t))
\, =\, J'_{lk_1} J'_{lk_2} Tr (
I_{lv,k_1v_1,k_2v_2} \rho'(t))\ee for all $v,v_1,v_2\in \{x,y,z\}$
and all corresponding trajectories $\rho(t)$ and $\rho'(t)$. Now
we apply to equation (\ref{p21}) statement $2.$ of Lemma
\ref{plemma2}, i. e. equation (\ref{p7}) with  $\bar{k}=k_1$ and
$\bar{l}=l$ to get: \be{p22}
  J^2_{lk_1} J_{lk_2}Tr ( I_{k_1v_1,k_2v_2} \rho(t))
\, =\, J^{'2}_{lk_1} J'_{lk_2} Tr (
I_{k_1v_1,k_2v_2} \rho'(t)).\ee On the other hand, we can  apply to
equation (\ref{p21}) again statement $2.$ of Lemma \ref{plemma2},
i. e.  equation (\ref{p7}) this time with $\bar{k}=k_2$ and
$\bar{l}=l$ to get:\be{p23}
  J_{lk_1} J^2_{lk_2}Tr ( I_{k_1v_1,k_2v_2}
\rho(t)) \, =\, J_{lk_1}' J^{'2}_{lk_2} Tr (
I_{k_1v_1,k_2v_2} \rho'(t)).\ee Since $J^2_{lk_1} J_{lk_2}= -
J^{'2}_{lk_1} J'_{lk_2}$ while $J_{lk_1} J^2_{lk_2}= J_{lk_1}'
J^{'2}_{lk_2}$ (by (\ref{p20})),   equations (\ref{p22}) and
(\ref{p23}) imply $Tr ( I_{k_1v_1,k_2v_2} \rho(t)) \, =0$ for
all $v_1,v_2\in \{x,y,z\}$ and all  trajectories $\rho(t)$. This
fact   contradicts the controllability assumption, thus equation
(\ref{p19}) holds.

  If
$J_{kl}=J'_{kl}$ for every pair $1\leq k<l \leq n$, from the
observability of the model, we must have $\rho_0=\rho_0'$, thus
equation  (\ref{puguali}) holds.

  On the other hand, if $J_{kl}'=-J_{kl}$ for every pair $1\leq k<l \leq
  n$, we argue as follows.
First, we prove, by induction on $1\leq r \leq n$ that: \be{p30}
Tr(I_{k_1v_1,...,k_rv_r}\rho(t))=
(-1)^{r-1}Tr(I_{k_1v_1,...,k_rv_r}\rho'(t)), \ \ \forall v_j\in
\{x,y,z\}, \ 1\leq k_1<\ldots <k_r\leq n, \ee and for all
corresponding trajectories $\rho(t)$ and $\rho'(t)$.

For $r=1$ the previous equation is equation (\ref{p2}), thus the
result holds for $r=1$. Assume that (\ref{p30}) holds for $1<r<n$,
and consider an arbitrary set of indices $1\leq k_2<\ldots
<k_{r+1}$. By the inductive assumption we have:
\[
Tr(I_{k_1v_1,...,k_rv_r}\rho(t))=
(-1)^{r-1}Tr(I_{k_1v_1,...,k_rv_r}\rho'(t)).
\]
Since the graph associated to the network  is connected, for each node
$k_j$, $j=1,\ldots,r$, there exists a path joining the node $k_j$
with  the node $k_{r+1}$. Let $\bar{j}$ be the index for which
this path is the shortest one, and denote by $l_1,\ldots,l_d$ the
intermediate nodes.  By the way we have chosen $\bar{j}$, it is
easy to see that
$\{k_1,\ldots,k_r\}\cap\{l_1,\ldots,l_d\}=\emptyset$. To fix
notations, we may assume without loss of generality (being all the
other cases the same) that $1\leq k_1<\ldots<k_r<l_1<\ldots
<l_d<k_{r+1}$. By applying statement $1.$ of Lemma \ref{plemma2},
equation (\ref{p6}) with $\bar{k}=\bar{j}$ and $\bar{l}=l_1$,
since $J_{\bar{j}l_1}=-J'_{\bar{j}l_1}$, we have:
\[
Tr(I_{k_1v_1,...,k_rv_r,l_1w_1}\rho(t))=
(-1)(-1)^{r-1}Tr(I_{k_1v_1,...,k_rv_r,l_1w_1}\rho'(t)),
\]
for any $w_1\in \{x,y,z\}$. By applying again statement $1.$ of
Lemma \ref{plemma2}, equation (\ref{p6})  another $d-1$-times with
$\bar{k}=l_i$ and $\bar{l}=l_{i+1}$, $i=1,\ldots,d-1$, and then
another  time with $\bar{k}=l_d$ and $\bar{l}=k_{r+1}$, we end up
with: \be{p31}
Tr(I_{k_1v_1,...,k_rv_r,l_1w_1,\ldots,l_dw_d,k_{r+1}v_{r+1}}\rho(t))=
(-1)^d
(-1)^{r}Tr(I_{k_1v_1,...,k_rv_r,l_1w_1,\ldots,l_dw_d,k_{r+1}v_{r+1}}\rho'(t)
).
\ee Now we apply to equation (\ref{p31}), statement $2.$ of  Lemma
\ref{plemma2}, i.e.
  equation (\ref{p7})  $d-1$-times with $\bar{k}=l_{i+1}$ and
  $\bar{l}=l_i$, $i=1,\ldots,d-1$, to get:
  \[
Tr(I_{k_1v_1,...,k_rv_r,l_dw_d,k_{r+1}v_{r+1}}\rho(t))=
(-1)^{d-1}(-1)^d
(-1)^{r}Tr(I_{k_1v_1,...,k_rv_r,l_dw_d,k_{r+1}v_{r+1}}\rho'(t)).
\]
Finally, by applying again  statement $2.$ of  Lemma
\ref{plemma2}, i.e.
  equation (\ref{p7}) with $\bar{k}=k_{r+1}$ and
  $\bar{l}=l_d$, we end up with:
  \[
Tr(I_{k_1v_1,...,k_rv_r,k_{r+1}v_{r+1}}\rho(t))=
(-1)^{r}Tr(I_{k_1v_1,...,k_rv_r,k_{r+1}v_{r+1}}\rho'(t)), \] as
desired. Thus equation (\ref{p30}) holds.

Now, denoting by $\rho_1$ and $\rho_2$ (resp. $\rho_1'$ and
$\rho_2'$)  the components of $\rho_0$ (resp. $\rho_0'$) in ${\cal
I}_o$, ${\cal I}_e$, by using (\ref{p30}), we have: \be{p32}
Tr\left((I_o(\rho_0- \rho_0')\right) =0, \ \ \
Tr\left((I_e(\rho_0+ \rho_0')\right) =0, \ee for all elements
$I_0\in {\cal I}_o$ and $I_e\in{\cal I}_e$. Equation (\ref{p32})
implies that the components of $\rho_0$ and $\rho_0'$ in ${\cal
I}_o$ coincide while the components in  ${\cal I}_e$ are opposite
to each other. Thus equation  (\ref{popposti}) holds.

\subsection{$(b) \ \Rightarrow \ (a)$}

 Let
$(\Sigma,\rho_0)$ and $(\Sigma',\rho'_0)$ be two   models, which
  are both controllable, with all the $\gamma_k$ and
$\gamma_k'$ different from each other. 

In the case where equation (\ref{puguali}) holds (i.e. same model and
same initial condition),  obviously 
that:
\[
\rho'(t)=\rho(t),
\]
for all $t\geq 0$. Thus the two models are equivalent.

Assume now that equation (\ref{popposti}) holds.  Thus, 
\[
A'=-A, \ \text{ and } \ B'_v=B_v \ \forall \ v\in \{x,y,z\}.
\]
and
\[
\rho_1=\rho_1' \ \text{ and } \ \rho_2=-\rho_2'. 
\]
We have:
  \be{C20} \dot
\rho(t)=\left[A+B_xu_x(t)+B_yu_y(t)+B_zu_z(t),\rho(t)\right], \ee
while \be{C21} \dot
\rho'(t)=\left[-A+B_xu_x(t)+B_yu_y(t)+B_zu_z(t),\rho'(t)\right].
\ee It is easily  verified   (cfr.  Lemmas \ref{plemma1} and \ref{plemma2})  that:
  \be{C22}
  \begin{array}{lll}
[B,I_{o}]\in  {\cal I}_{o}, &  [A,I_{o}]\in  {\cal I}_{e}, &
\forall I_o \in {\cal I}_o, \\ {[}B,I_{e}]\in  {\cal I}_{e}, &
[A,I_{e}]\in  {\cal I}_{o},  &
  \forall I_e \in {\cal I}_e. \end{array} \ee
  Thus, we can 
write the differential equations for $\rho_1(t)$ and $\rho_2(t)$
as: \be{C23} \begin{array}{lcl}
  \dot \rho_1(t) &=
  &[B_xu_x(t)+B_yu_y(t)+B_zu_z(t),\rho_1(t)]+[A,\rho_2(t)]\\
  \dot \rho_2(t) & =& 
[A,\rho_1(t)]+[B_xu_x(t)+B_yu_y(t)+B_zu_z(t),\rho_2(t)]
\end{array} \ee
and similarly the differential equation for $\rho_1'(t)$ and
$\rho_2'(t)$ as: \be{C24} \begin{array}{lcl}
  \dot \rho_1'(t) &=
  &[B_xu_x(t)+B_yu_y(t)+B_zu_z(t),\rho_1'(t)]-[A,\rho_2'(t)]\\
  \dot \rho_2'(t) & =& -[A,\rho_1'(t)]+
[B_xu_x(t)+B_yu_y(t)+B_zu_z(t),\rho_2'(t)]
\end{array} \ee
Combining equations (\ref{C23}) and (\ref{C24}),  we obtain a
differential equation for $\rho_1(t)-\rho_1'(t)$ and for
$\rho_2(t)+\rho_2'(t)$. In particular, we have \be{C25}
\begin{array}{lcl}
  \dot \rho_1(t)- \dot \rho_1'(t) &=
  &[B_xu_x(t)+B_yu_y(t)+B_zu_z(t),\rho_1(t)-
\rho_1'(t)]+[A,\rho_2(t)+  \rho_2'(t)]\\
  \dot \rho_2(t)+ \dot \rho_2'(t) & =&
   [A,\rho_1(t)-
   \rho_1'(t)]+[B_xu_x(t)+B_yu_y(t)+B_zu_z(t),\rho_2(t)+
   \rho_2'(t)]
\end{array} \ee
 From equations (\ref{C25}) it follows that if
$\rho_1(0)=\rho_1'(0)$ and
  $\rho_2(0)=-\rho_2'(0)$ then $\rho_1(t)=\rho_1'(t)$ and
  $\rho_2(t)=-\rho_2'(t)$, for every $t$ and for every
controls $u_x(t)$, $u_y(t)$, and $u_z(t)$. In
particular, since $Tr(S^{TOT}_v\rho(t))=Tr(S^{TOT}_v\rho_1(t))$
for all $v\in \{x,y,z\}$ and $\rho_1(t) \equiv \rho_1'(t)$ the two
models are equivalent.

\section{Conclusions}

  In this paper,  we have investigated methods of dynamic parameter
identification for networks of spin $\frac{1}{2}$ particles. We have
shown that by driving the network with an appropriate electro-magnetic
field and measuring the total magnetization in a given (arbitrary)
direction it is possible to identify the parameters.
Moreover, if the initial state is not known, it is possible to
obtain combined information about the initial state and the parameter
values. We have assumed that all the gyromagnetic ratios of the spins
are different or that it is possible to address each spin separately. In
the opposite case,  where all the gyromagnetic ratios are the same,  the
unitary evolution $X(t)$, solution of Schr\"odinger operator equation,
  has the form $X(t)=e^{At}\Phi(t)$ where $\Phi(t)$ depends only on the
controls $u_x, u_y, u_z$ and  $A$ is defined in (\ref{dinamica}),
(\ref{dinamica1}). In this case,  we have that  $A$ 
commutes with $\Phi$ and $S_{v}^{TOT}$,
$v=x,y,z$, and therefore  the 
output $Tr(S_v^{TOT}e^{At}\Phi(t)\rho_0 \Phi^*(t)e^{-At})$
is equal to $Tr(e^{-At}S_v^{TOT}e^{At}\Phi(t)\rho_0 \Phi^*(t))=
Tr(S_v^{TOT}\Phi(t)\rho_0 \Phi^*(t))$. The output is 
 therefore independent of
$A$. This implies that it is not possible to identify the parameters in
$A$ by a reading of the total magnetization.

In our approach, the system theoretic concepts of
controllability and observability as well as previously  known results on
the controllability of spin networks have played an important role. This
is usually the case in the theory of parameter identification and we
believe this approach will be useful for other classes of quantum
systems. Extensions of the results presented here are possible and
will be object of further research. For example, the hypothesis of
controllability of the models can be weakened. If a spin network is not
controllable and has different gyromagnetic ratios the 
associated graph has several connected components. The dynamical Lie
Algebra associated to the system  is the direct sum of
Lie Algebras isomorphic to $su(2^{n_j})$ where $n_j$ is the number of
nodes (spins) in the $j-$th component \cite{confraLAA}. 
Another important research
problem is the actual design of control algorithms for parameter
identification for  which the research presented here is a preliminary
step.

\section*{Appendix A: Proof of Proposition \ref{prop1}}
\bpr
It is clear that (b) implies (a), since
$iS^{TOT}_{v}\in \cal V$ ($iS'^{TOT}_{v}\in {\cal V}'$) for
$v=x,\,y,\, z.$ We will prove the converse implication by
induction on the depth $s=\sum_{i=1}^r k_i$ in (\ref{mat}) of the
matrix $F\in \cal V$. If $s=0$ then $F=iS^{TOT}_{v}$ for $v=x,\,
y$ or $z$, thus equation (\ref{tra}) holds by definition of
equivalence. Assume that equation (\ref{tra}) holds for matrices
in $\cal V$ of depth $\leq s$ and let $F\in \cal V$ with  depth
equal $s+1$. Then
\[
F=ad_{B_{\bar{i}}} G, \ \ \text{ with } G\in \cal V, \] and the depth
of $G$ is equal to $s$. Assume, by contradiction,  that there exist
control functions $u_x(\cdot)$, $u_y(\cdot)$, and $u_z(\cdot)$,
and $\bar{t}\geq 0$ such that equation (\ref{tra}) does not hold,
i.e. \be{contra} Tr(F\rho(\bar{t})) \, \neq \,
Tr(F'\rho'(\bar{t})).\ee On the other hand, since by the inductive
assumption, equation (\ref{tra}) holds for the matrix $G$, we
have:
\[
\frac{d}{dt} Tr(G\rho(t)) \, =\,\frac{d}{dt}
  Tr(G'\rho'(t)),
\]
for all $t\geq 0$. This implies: \be{eq1} Tr( [G,B_0] \rho(t)) +
  Tr([G,B_1]\rho(t))u_x(t) +
Tr([G,B_2]\rho(t))u_y(t) + \] \[ Tr([G,B_3]\rho(t))u_z(t) \, = \,
Tr( [G',B_0'] \rho'(t)) + Tr([G',B_1']\rho'(t))u_x(t) + \]
\[ +
Tr([G',B_2']\rho'(t))u_y(t) + Tr([G',B_3']\rho'(t))u_z(t) \ee
Define:
\[
u^0_v(t)\, =\, \left\{ \begin{array}{cl}
                u_v(t)  & \text{ for }t< \bar{t} \\
                0       & \text{ for } t\geq \bar{t}.
                \end{array} \right. \]
Then, clearly the trajectories $\rho(t)$ and $\rho'(t)$
corresponding to the two sets  of controls $u_v(\cdot)$ and
$u^0_v(\cdot)$ are equal up to time $\bar{t}$.  Thus evaluating
(\ref{eq1}) at $t=\bar{t}$,  using controls $u^0_v$,  we get:
\be{iolo}
   Tr( [G,B_0] \rho(\bar{t}))\, =\,
  Tr( [G',B_0'] \rho'(\bar{t})),
\ee
  which contradicts (\ref{contra})  if $\bar i=0$.
Assume  $\bar{i}\neq 0$. First notice that,  by repeating the same
argument as above for a generic $t\geq 0$,  the
previous equality (\ref{iolo}) must hold for all $t \geq 0$. To
get a contradiction we use the control functions $u^{\bar{i}}_v$
given by\footnote{$\delta_{1x}\equiv \delta_{2y} \equiv \delta_{3z}
\equiv 1$, $\delta_{\bar i v} \equiv 0$ otherwise.}:
\[
u^{\bar{i}}_v(t)\, =\, \left\{ \begin{array}{cl}
                u_v(t)  & \text{ for }t < \bar{t} \\
                \delta_{\bar{i},v}       & \text{ for } t\geq\bar{t}.
                \end{array} \right. \]
Again the trajectories corresponding to the two set of controls
$u_v(\cdot)$ and $u^{\bar{i}}_v(\cdot)$ are equal up to time
$\bar{t}$, thus evaluating (\ref{eq1}) at $t=\bar{t}$ using
controls $u^{\bar{i}}_v$ we get:
\[
  Tr( [G,B_0] \rho(\bar{t})) + 
Tr([G,B_{\bar{i}}]\rho(\bar{t}))u_{\bar{i}}(\bar{t})
\,
   =
  Tr( [G',B'_0] \rho'(\bar{t}))+ Tr([G',B'_{\bar{i}}]
  \rho'(\bar{t}))u_{\bar{i}}(\bar{t})
  \]
  which,  since the first terms are equal
as observed before, contradicts (\ref{contra}),
  and ends the proof.
  \epr

\section*{Appendix B: Proofs of the
preliminary results in section \ref{preli}}

\subsection*{Proof of Proposition \ref{prop2}}
\bpr It is not difficult to see that, for all $l\geq 0$, if $
F^l=\sum_{k=1}^n  \gamma^l_k I_{kv}\in {\cal{V}},$ then the
corresponding matrix in $\cal V'$ is ${F'}^l=\sum_{k=1}^{n'}
{\gamma'}^l_k I_{kv}$. By Proposition \ref{prop1}, it holds that:
\[
Tr( F^l\rho(t)) \,= \, Tr({F'}^l\rho'(t)).
\]
for all $t\geq 0$ and all possible trajectories. Thus we have:
\be{e2} \sum_{k=1}^n \gamma^l_k Tr(I_{kv}\rho(t))\, = \,
\sum_{k=1}^{n'} \gamma^{'l}_k Tr(I_{kv}\rho'(t)). \ee Fix $v$  and
let $\alpha_{k}(t)=Tr(I_{kv}\rho(t))$ and
$\alpha'_{k}(t)=Tr(I_{kv}\rho'(t))$, then we  rewrite equation
(\ref{e2}) as: \be{e3} \sum_{k=1}^n \gamma^l_k \alpha_{k}(t) -
\sum_{k=1}^{n'} \gamma'^l_k \alpha'_{k}(t) \, =\, 0 \ee The matrix
$M\in \r^{(n+n')\times (n+n')}$ given by:
\[
M\, =\, \left( \begin{array}{cccccc}
                   1&\ldots& 1& 1& \ldots& 1\\
      \gamma_{1}&\ldots& \gamma_{n}& \gamma'_{1}& \ldots& \gamma'_{n'}\\
      \gamma^{2}_{1}&\ldots& \gamma^{2}_{n}& \gamma'^{2}_{1}& \ldots&
      \gamma'^{2}_{n'}\\
       & & & & & \\
       & & & & & \\
       \gamma^{n+n'}_{1}&\ldots& \gamma^{n+n'}_{n}& \gamma'^{n+n'}_{1}& 
\ldots&
      \gamma'^{n+n'}_{n'}
      \end{array} \right),
      \]
is a Vandermonde type of matrix. Notice that the coefficients
$\gamma_{k}$, $k=1,\ldots,n$ and also $\gamma'_{k}$,
$k=1,\ldots,n'$, are all different. Moreover, the coefficients
$\alpha_{k}(t)$ and $\alpha'_{k}(t)$ are not identically zero. In
fact, if $\alpha_k(t)$ was identically zero we would have
$Tr(A\rho(t))=0$, for every $A \in su(2^n)$ (by the
controllability assumption) which would imply $\rho_{0}$ equal to a
multiple of the identity matrix which we have  excluded.
Thus, from equation
(\ref{e3}), we conclude that there exist two indices $\bar{k}$ and
$\pi({\bar{k})}$ such that
\[
\gamma_{\bar{k}}=\gamma'_{\pi({\bar{k}})}.
\]
We can  rewrite equation (\ref{e3}), as \be{e4}
\gamma_{\bar{k}}^{l} \left(
\alpha_{\bar{k}}(t)-\alpha'_{\pi(\bar{k})}(t) \right) +
\sum_{k=1,k\neq\bar{k}}^n \gamma^l_k \alpha_{k}(t) -
\sum_{k=1,k\neq\pi(\bar{k})}^{n'} \gamma'^l_k \alpha'_{k}(t)  \,
=\, 0 \ee Now we can  repeat the same argument
and, unless $n=1$ or $n'=1$, we will find two more indices
$\bar{j}$ and $\pi(\bar{j})$ whose corresponding values of
$\gamma_{\bar{j}}$ and $\gamma'_{\pi(\bar{j})}$ are equal. We may
assume without loss of generality that $n' \geq n$ and repeat this
procedure $n$-times. Thus we find a permutation  $\pi$ from the set
$\{1,\ldots,n\}$ to the set $\{1,\ldots,n'\}$ and we rewrite
equation (\ref{e4}) as: \be{e5} \sum_{k=1}^n \gamma^l_k \left(
\alpha_{k}(t) - \alpha'_{\pi(k)}(t) \right) -
\sum_{\begin{array}{l}k\neq \pi(j)\\j=1,\ldots,n
\end{array}} \gamma'^l_k \alpha'_{k}(t) \, =\, 0. \ee Now, we can
apply again the same argument, using the Vandermonde matrix $N$
constructed with all the coefficients $\gamma_{k}$ and the
coefficients $\gamma'_{k}$ for those indices that are not in the
image of $\pi$. Since the coefficients  $\alpha'_{k}(t)$ are not
identically zero, we can  conclude that all the coefficients
$\gamma'_{k}$,  for those indices that are not in the image of $\pi$, 
must be zero. Thus,  in particular $n=n'$, the map $\pi$ is a
permutation and :
\[
\gamma_{{k}}=\gamma'_{\pi({{k}})}, \ \
\alpha_{{k}}(t)-\alpha'_{\pi({k})}(t) = 0, \qquad \ \forall t \geq 0,
\]
which concludes the proof.
\epr

\subsection*{Proof of Lemma \ref{lemma2}}
\bpr We will prove
the result by induction on the depth $r$ of $K$ and $K'$. If $r=0$, the result follows
from  Lemma \ref{lemma1}. Now assume that for every
pair $K$ and $K'$ of depth $r$,  and every pair of matrices $W$
and $W'$,  (\ref{C3}) implies (\ref{C4}). From the inductive
assumption we have \be{C5} Tr([W,K]\rho(t))=Tr([W',K']\rho'(t)),
\ee and \be{C6} Tr([W,B_j]\rho(t))=Tr([W',B_j']\rho'(t)). \ee
Applying the inductive assumption with $W$ ($W'$) replaced by
$[W,K]$ and $[W,B_j]$ ($[W',K']$ and $[W',B_j']$) we obtain
\be{C7} Tr([[W,K],B_j]\rho(t))=Tr([[W',K'],B_j']\rho'(t)), \ee
\be{C8} Tr([[W,B_j],K]\rho(t))=Tr([[W',B_j'],K']\rho'(t)). \ee
Combining (\ref{C7}) and (\ref{C8}) using the Jacobi identity, we
obtain \be{C9} Tr([W,[K,B_j]]\rho(t))= Tr([W',[K',B_j']]\rho'(t)),
\ee which proves the lemma. \epr

\subsection*{Proof of Lemma \ref{lemma4}}
\bpr For any $l\geq 0$  let
$F_l:=\sum_{k=1}^n \gamma_k^l I_{kv}\in \cal V$; then its
corresponding matrix in $\cal V'$ is $F_l':=\sum_{k=1}^n
\gamma_{\pi(k)}^l I_{\pi(k)v}$. By  applying Lemma \ref{lemma2},
with $K=F_l$ and $K'=F'_l$, we obtain \be{C14} \sum_{k=1}^n
\gamma_k^l Tr([I_{kv},W] \rho(t))= \sum_{k=1}^n
\gamma_{\pi{(k)}}^l Tr([I_{\pi{(k)}v},W'] \rho'(t)). \ee
  Using the fact that $\gamma_k=\gamma_{\pi{(k)}}$, we can 
rewrite equation (\ref{C14})
  as
  \[
  \sum_{k=1}^n \gamma_k^l\left( Tr([I_{kv},W] \rho(t))- 
Tr([I_{\pi{(k)}v},W']
  \rho'(t))\right) = 0.
  \]
  Since the coefficients $\gamma_k$ are all different and the
  previous equality holds for every $l\geq 0$, using
a Vandermonde determinant type of argument, we obtain \be{C15}
Tr([I_{kv},W] \rho(t))-Tr([I_{\pi(k)v},W'] \rho'(t))=0, \ee as
desired. \epr

\section*{Appendix C: Proofs of the lemmas in Section \ref{complet}}

\subsection*{Proof of Lemma \ref{plemma1}}

\bpr

1. This fact follows easily by applying Lemma \ref{lemma4} with
$W=W'=I_{k_1v_{1},...,k_rv_{r}}$, since, if $v_j\neq w_j$, it
holds: \be{p8} \left[ I_{k_1v_{1},...,k_rv_{r}}, I_{k_jw_j}
\right] = i I_{k_1v_{1},\ldots,k_j[v_jw_j],\ldots,k_rv_{r}}. \ee
Here we have used the notation $[v_jw_j]=v$ if
$[\sigma_{v_j},\sigma_{w_j}]=\pm i \sigma_v$, and agreed  to
multiply (\ref{p8}) by $-1$ if the minus sign appears.

2. By applying Lemma \ref{lemma1} to equation (\ref{p3}), we have:
\be{p9} Tr \left( \left[ A, I_{k_1v_{1},...,k_rv_{r}}\right]
\rho(t) \right) = Tr \left( \left[ A',
I_{k_1v_{1},...,k_rv_{r}}\right] \rho'(t) \right). \ee Now, we
apply Lemma \ref{lemma4} to the previous equation to get: \be{p10}
  Tr \left( \left[ \left[ A, I_{k_1v_{1},...,k_rv_{r}}\right],
  I_{\bar{k}v_{\bar k}} \right]
\rho(t) \right) =   Tr \left( \left[ \left[ A',
I_{k_1v_{1},...,k_rv_{r}}\right],
  I_{\bar{k}v_{\bar k}} \right]
\rho'(t) \right). \ee Using the Jacobi identity, we have: \be{p11}
\left[ \left[ A, I_{k_1v_{1},...,k_rv_{r}}\right],
  I_{\bar{k}v_{\bar k}} \right] =
  \left[A, \left[  I_{k_1v_{1},...,k_rv_{r}},
  I_{\bar{k}v_{\bar k}} \right]\right] +
\left[I_{k_1v_{1},...,k_rv_{r}}, \left[ A,
  I_{\bar{k}v_{\bar k}} \right]\right]
  \ee
We have  \be{p12} \left[  I_{k_1v_{1},...,k_rv_{r}},
  I_{\bar{k}v_{\bar k}} \right]= \left\{ \begin{array}{cl}
                 0  & \text{if } \bar{k}\not\in
                 \{k_1,\ldots,k_r\}\\
                 0   & \text{if } \exists j \text{ with }\bar{k}=k_j 
\text{ and }
                    v_{\bar{k}}=v_j \\
                 i I_{k_1v_{1},\ldots,k_j[v_jv_{\bar{k}}],\ldots,k_rv_{r}}
                & \text{if } \exists j \text{ with }\bar{k}=k_j \text
{ and }
                    v_{\bar{k}}\neq v_j.
                     \end{array} \right.
\ee Using the fact that (\ref{p9}) holds for any choice of values
$v_k$, and (\ref{p12}) we get:
\[
Tr \left(\left[A, \left[  I_{k_1v_{1},...,k_rv_{r}},
  I_{\bar{k}v_{\bar k}} \right]\right]\rho(t) \right)
= Tr \left(\left[A', \left[  I_{k_1v_{1},...,k_rv_{r}},
  I_{\bar{k}v_{\bar k}} \right]\right]\rho'(t) \right).
  \]
  Thus combining the previous equality with (\ref{p10}) and
  (\ref{p11}) we get:
  \be{p13}
Tr \left( \left[ \left[ A, I_{\bar{k}v_{\bar k}} \right],
I_{k_1v_{1},...,k_rv_{r}} \right] \rho(t) \right) = Tr \left(
\left[ \left[ A', I_{\bar{k}v_{\bar k}} \right],
I_{k_1v_{1},...,k_rv_{r}} \right] \rho'(t) \right). \ee Notice
that equation (\ref{p13}) is of the same type as  equation
(\ref{p9}); it is enough to replace $A$ with
$[A,I_{\bar{k}v_{\bar{k}}}]$ (resp. $A'$ with
$[A',I_{\bar{k}v_{\bar{k}}}]$). Thus by applying first Lemma
\ref{lemma4} and then the Jacobi identity we get: \be{p14} Tr
\left( \left[ [A,I_{\bar{k}v_{\bar{k}}}], \left[
I_{k_1v_{1},...,k_rv_{r}},
  I_{\bar{l}v_{\bar l}} \right]\right] \rho(t) \right) +
  Tr \left( \left[ I_{k_1v_{1},...,k_rv_{r}},
\left[ [A,I_{\bar{k}v_{\bar{k}}}], I_{\bar{l}v_{\bar l}}\right]
\right] \rho(t) \right) =
\]
\[ =
  Tr
\left( \left[ [A',I_{\bar{k}v_{\bar{k}}}], \left[
I_{k_1v_{1},...,k_rv_{r}},
  I_{\bar{l}v_{\bar l}} \right]\right] \rho'(t) \right) +
  Tr \left( \left[ I_{k_1v_{1},...,k_rv_{r}},
\left[ [A',I_{\bar{k}v_{\bar{k}}}], I_{\bar{l}v_{\bar l}}\right]
\right] \rho'(t) \right). \ee On the other hand, using (\ref{p12})
and (\ref{p13}), we get:
\[  Tr
\left( \left[ [A,I_{\bar{k}v_{\bar{k}}}], \left[
I_{k_1v_{1},...,k_rv_{r}},
  I_{\bar{l}v_{\bar l}} \right]\right] \rho(t) \right)=
  Tr
\left( \left[ [A',I_{\bar{k}v_{\bar{k}}}], \left[
I_{k_1v_{1},...,k_rv_{r}},
  I_{\bar{l}v_{\bar l}} \right]\right] \rho'(t) \right).
  \]
  Thus:
  \[
  Tr \left( \left[ I_{k_1v_{1},...,k_rv_{r}},
\left[ [A,I_{\bar{k}v_{\bar{k}}}], I_{\bar{l}v_{\bar l}}\right]
\right] \rho(t) \right)=  Tr \left( \left[
I_{k_1v_{1},...,k_rv_{r}}, \left[ [A',I_{\bar{k}v_{\bar{k}}}],
I_{\bar{l}v_{\bar l}}\right] \right] \rho'(t) \right);
\]
which implies (\ref{p4}), as desired. \epr

\subsection*{Proof of Lemma \ref{plemma2}}

\bpr Both statements are a consequence of Lemma \ref{plemma1}
(equation (\ref{p4})). First notice that, again by Lemma
\ref{plemma1}, it is enough to prove (\ref{p6}) and (\ref{p7}) for
a particular choice of $\{v_j\}$ and $\bar{v}$. We have, for
$\bar{l}>\bar{k}$, \be{p15} \left[
iI_{\bar{l}z},\left[iI_{\bar{k}x}, A\right]\right] =
-J_{\bar{k}\bar{l}}i I_{\bar{k}z,\bar{l}x}. \ee

1. By applying Lemma \ref{plemma1} (equation (\ref{p4})) to
(\ref{p5}) and using (\ref{p15}) we get: \be{p16} \alpha
  Tr\left(\left[-J_{\bar{k}\bar{l}}i I_{\bar{k}z,\bar{l}x},  
I_{k_1v_1,...,k_r
  v_r}\right]
\rho(t)\right)=
  \alpha
  Tr\left(\left[-J'_{\bar{k}\bar{l}}i I_{\bar{k}z,\bar{l}x},  
I_{k_1v_1,...,k_r
  v_r}\right]
\rho'(t)\right). \ee We may assume, without loss of generality,
that $\bar{k}=k_j$ and $v_j=x$. In this case we have:
\[
-J_{\bar{k}\bar{l}} \left[ I_{\bar{k}z,\bar{l}x},
I_{k_1v_1,...,k_r
  v_r}\right]=J_{\bar{k}\bar{l}} i
  I_{k_1v_1,\ldots,k_jy,\ldots,k_rv_r,\bar{l}x}.
  \]
  Combining the previous equality with (\ref{p16}), equation
  (\ref{p6}) follows easily.

  2. Using the same procedure, we end up again with equation
  (\ref{p16}), but now both indices $\bar{k}$ and $\bar{l}$ are in
  $\{k_1,\ldots,k_r\}$. Assume, for example that $k_1=\bar{k}$ and
  $k_2=\bar{l}$, and take $v_{k_1}=v_{k_2}=x$, then equation
  (\ref{p7}) follows since it holds:
  \[
\left[ I_{k_1z,k_2x}, I_{k_1x,k_2x,\ldots,k_r
  v_r}\right]=1/4
  I_{k_1y,k_3v_3,\ldots, k_rv_r}.
  \]
\epr

\end{document}